\documentclass[aps,preprint]{revtex4}
\usepackage{amssymb}
\usepackage{latexsym}
\usepackage{amsfonts}
\usepackage{graphicx}
\usepackage{epsfig}
\usepackage{amsmath}
\usepackage{float}  
\usepackage{verbatim}

\begin{document}

\title{Statistical Regularities of Equity Market Activity}

\author{Fengzhong Wang,$^1$ Kazuko Yamasaki,$^{1,2}$ Shlomo
Havlin,$^{1,3}$ and H. Eugene Stanley$^1$}

\affiliation{$^1$Center for Polymer Studies and Department of Physics,
Boston University, Boston, MA 02215 USA\\$^2$Department of
Environmental Sciences, Tokyo University of Information Sciences,
Chiba 265-8501,Japan\\$^3$Minerva Center and Department of Physics,
Bar-Ilan University, Ramat-Gan 52900, Israel}

\date{14 November 2009 ~~~ wyhs.tex}

\begin{abstract}

Equity activity is an essential topic for financial market studies. To
explore its statistical regularities, we comprehensively examine the
trading value, a measure of the equity activity, of the $3314$
most-traded stocks in the U.S. equity market and find that (i) the
trading values follow a log-normal distribution; (ii) the standard
deviation of the growth rate of the trading value obeys a power-law
with the initial trading value, and the power-law exponent
$\beta=0.14$. Remarkably, both features hold for a wide range of
sampling intervals, from $5$ minutes to $20$ trading days. Further, we
show that all the $3314$ stocks have long-term correlations, and their
Hurst exponents $H$ follow a normal distribution. Furthermore, we find
that the Hurst exponent depends on the size of the company. We also
show that the relation between the scaling in the growth rate and the
long-term correlation is consistent with $\beta=1-H$, similar to that
found recently on human interaction activity by Rybski and
collaborators.

\end{abstract}


\maketitle

\section{Introduction}

As a typical complex system, the financial markets attracts many
researchers in both economics and physics
\cite{Kondor99,Bouchaud00,Mantegna00,Johnson03,Cizeau97,Liu97,Liu99,Harris86,Admati88,Plerou01,Plerou07,Lux00,Giardina01,Yamasaki05,Wang06,Wang07,Wang09,Weber07,Ivanov04,Eisler06A,Eisler06B,Eisler08}. It
has been extensively studied over one hundred years, especially in
recent few decades, since huge financial databases became available
due to the development of electronic trading and data storing. A key
issue of these studies is the dynamics of the equity market, including
both of the price movement and market activity. Several stylized facts
have been found for the equity price movement, such as (i) the
distribution of the stock price changes (``return'') has a power-law
tail and (ii) the absolute value of price change (``volatility'') is
long-term power-law correlated
\cite{Kondor99,Bouchaud00,Mantegna00,Johnson03,Cizeau97,Liu97,Liu99,Harris86,Admati88,Lux00,Giardina01}. These
measures have been well studied, for example, a recent approach called
return interval analysis has been developed to comprehensively study
the temporal structure in the volatility time series
\cite{Yamasaki05,Wang06,Wang07,Wang09}.

The market activity was also studied by many researchers. Plerou et
al. investigated the market activity and found that the number of
trades displays long-term power-law correlations and the trading
volume follows a L\'{e}vy-stable distribution
\cite{Plerou01,Plerou07}. Ivanov et al. studied the inter-trade time
and showed multiscaling behavior in its distribution
\cite{Ivanov04}. Eisler and Kert\'{e}sz analyzed the fluctuation in
the trading values and found a certain scaling law
\cite{Eisler06A,Eisler06B,Eisler08}. Moreover, the activity in many
other economic and social systems has been studied
\cite{Gibrat31,Zipf32,Sutton97,Gabaix99,Stanley96,Rybski09}. For
example, recently Rybski et al. studied the dynamics of human
interaction activity and showed the connection between the long-term
correlations in the activity and scaling in the activity growth
\cite{Rybski09}. It is important to comprehensively examine the
dynamics of the equity activity and test the relation between the
correlation and the scaling in the growth rate, which may help to
better understand financial markets. In addition, the comparison
between the financial markets and other complex systems may shed light
on revealing the underlined mechanisms of the complex systems. For
this purpose, we study here the equity activity of the U.S. stock
market, a representative example of the world financial markets.

The paper is organized as follows: In section II we introduce the
database and the variable of trading value which characterizes the
equity activity, and demonstrate the intraday pattern for the trading
value. In section III we investigate the distribution of the trading
values and find that the distribution follows a log-normal
function. We also find a power-law relation between the standard
deviation of the growth rate of the trading value and the initial
trading value, which holds for a wide range of sampling
intervals. Section IV deals with the long-term correlations in the
time series of the market activity, which is characterized by the
Hurst exponent $H$. We show that the Hurst exponents of the stocks
follow a normal distribution with $H=0.75\pm0.09$. In Section V we
discuss the relation between the scaling in the growth rate and the
long-term correlations, and summarize our findings.

\section{Data Analyzed}

In this paper we analyze the Trades And Quotes (TAQ) database from the
New York Stock Exchange (NYSE), which records every transaction for
all securities in the U.S. equity market. The period studied is from
January 2, 2001 to December 31, 2002 (in total $500$ trading
days). However, the number of trading days varies with the stock. To
have enough records for analyzing, we only consider the stocks that
were traded at least $480$ days. In total we have $3314$ stocks which
include $1.42 \times 10^9$ records. In addition, these stocks have
quite different trading frequencies, from $6$ times per day to $6
\times 10^4$ times per day. To study different stocks on the same
footing, we adopt several typical sampling intervals $\Delta T$ in our
analysis, including $\Delta T= 5$-min, $30$-min, $1$-day, $5$-day ($1$
trading week), and $20$-day (roughly $1$ trading month). Note that $1$
trading day has $390$ minutes in the U.S. market. Thus the range of
these sampling intervals is over $3$ order of magnitudes.

\begin{figure*}
\begin{center}
   \includegraphics[width=\textwidth, angle = 0]{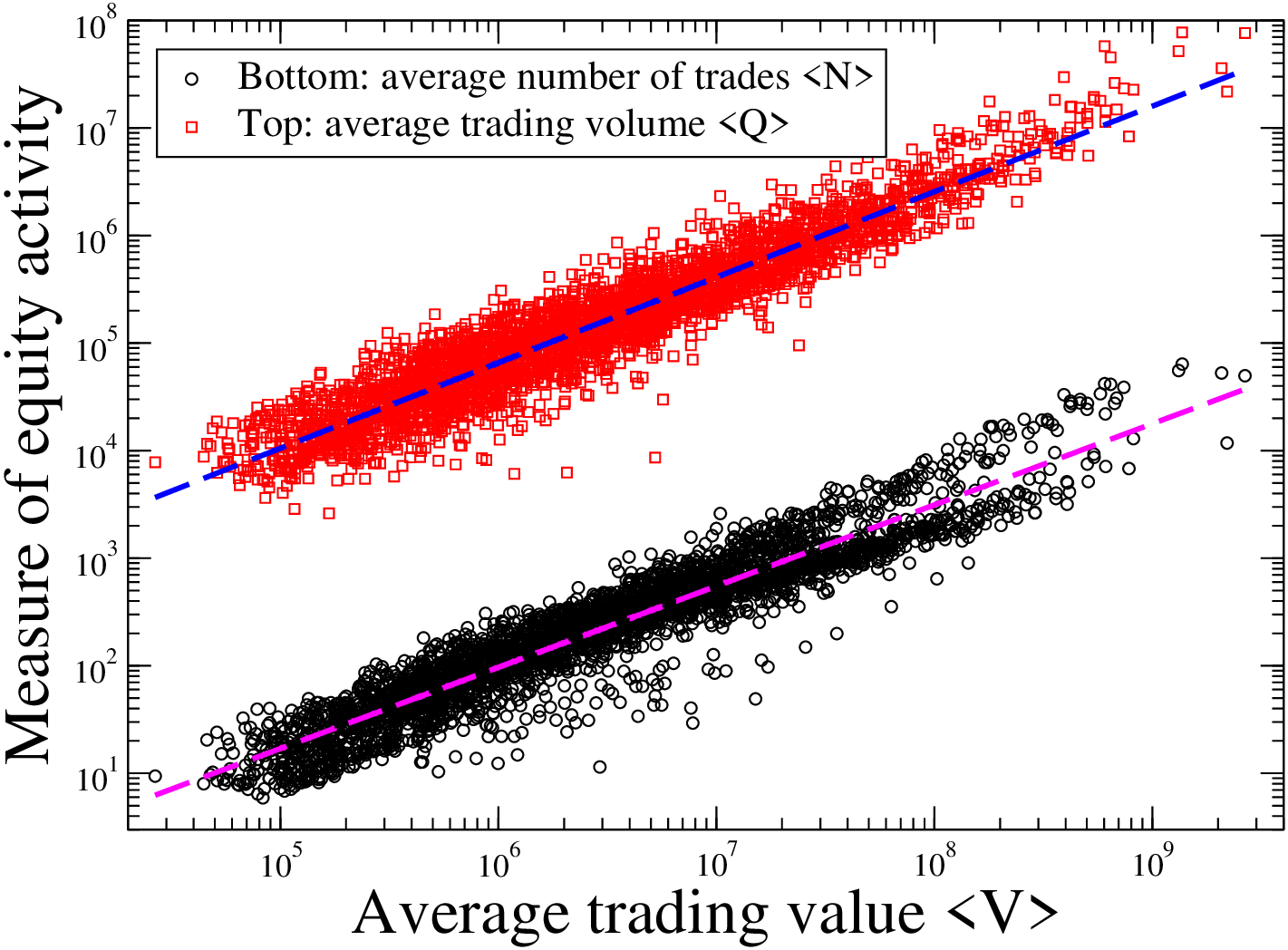}
\end{center}
\caption{(Color online) Relation between the trading value $V$ and two
other measures of the equity activity, number of trades $N$ and
trading volume $Q$. Each data point represents the daily average of
one stock in the 2001-02 period, and in total $3314$ stocks are
plotted. Clearly the three measures are strongly correlated. To
further show the tendency, we fit both cases with the power-law, as
shown by the dashed lines. The corresponding power-law exponent is
$0.76$ for the number of trades and $0.80$ for the trading volume. }
\label{Fig1}
\end{figure*}

The trading activity of an equity can be characterized by several
measures, such as the number of trades $N$, trading volume $Q$ (number
of shares traded), and trading value $V$ (amount of money traded) in a
certain time interval $\Delta T$. To select an activity measure, we
first need to test the relation among $V$, $N$, and $Q$. Without loss
of generality, we adopt $\Delta T= 1$-day. In Fig \ref{Fig1} we plot
the average number of trades $\langle N \rangle$ vs. the average
trading value $\langle V \rangle$ and the average trading volume
$\langle Q \rangle$ vs. the average trading value $\langle V
\rangle$. In this paper $\langle ... \rangle$ stands for the average
over the whole data set. Both cases show straight-line tendencies in
the log-log scale, suggesting there are strong dependence between
them. The correlation is $0.81$ between $\langle N \rangle$ and
$\langle V \rangle$, and $0.86$ between $\langle Q \rangle$ and
$\langle V \rangle$. Such strong correlations indicate that the three
variables have similar features. In addition, the unit of the trading
value (dollar) is the same for all stocks, but the unit of number of
trades (times of stock) and trading volume (shares of stock) are all
specific to a certain stock, and any two stocks are not the exact same
financial asset. Thus we choose the trading value $V$ as the measure
of the market activity.

In contrast to daily equity data, the intraday data are known to show
specific patterns for measures such as volatility
\cite{Liu99,Harris86,Admati88,Wang06}, due to different behaviors of
traders during a trading day. For example, the market is very active
immediately after the opening \cite{Admati88}, due to information
arriving while the market is closed. A question naturally arises. Is
there any intraday pattern in the trading activity? To test this, we
investigate the daily trend of the trading value for all the $3314$
stocks. For one stock, the intraday pattern $A(s)$ is defined as

\begin{equation}
A(s)\equiv \frac{\langle V \rangle_s}{\langle V \rangle},
\end{equation}

\begin{figure*}
\begin{center}
   \includegraphics[width=\textwidth, angle = 0]{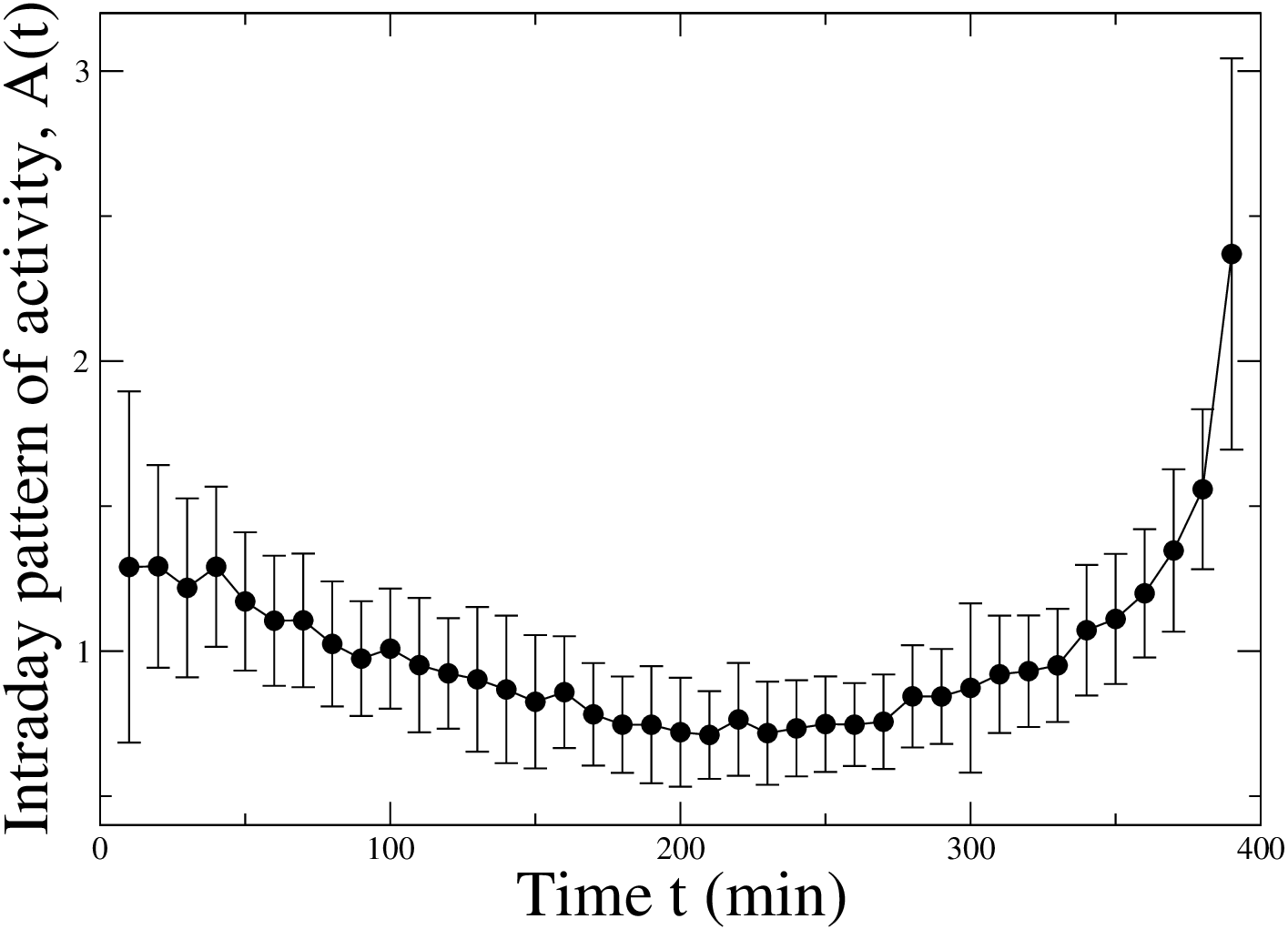}
\end{center}
\caption{(Color online) Intraday pattern of the trading values. For
each stock, we obtain the intraday pattern by computing the average
trading value over all trading days for each 10-min time interval, and
normalize it with the average activity over all 10-min intervals of
this stock. To demonstrate the pattern of all equities in the market,
we plot the average and standard deviation (as error bar) over the
patterns of all the $3314$ stocks against the time $t$ in a trading
day. Clearly we can see that the market activity is not uniform during
a trading day. In particular it is relatively intensive in the opening
and closing hours, and quiet around noon. Moreover, the differences
are significantly large than the error bars, suggesting that this
U-shape pattern persists for the entire market. Note that the error
bars are larger in the opening and closing hours, which is consistent
with the intensive activity in these hours. }
\label{Fig2}
\end{figure*}

Here $\langle V \rangle_s$ is the average trading value at a specific
moment $s$ of a trading day, and $\langle V \rangle$ is the average
trading value over all records. To show the tendency over the whole
market, we plot the average of $A(s)$ of all the $3314$ stocks and its
standard deviation (as error bar) in Fig. \ref{Fig2}. Clearly there is
no uniform pattern during a trading day. The pattern has a minimum
around noon ($s=200$ min), which is consistent with the intraday
pattern found for volatility
\cite{Liu99,Harris86,Admati88,Wang06}. However, there is a pronounced
peak at the closing hours and relative high values at the opening
hours, which are opposite to that of volatility where in the opening
hours it is higher compared to the closing hours. The volatility is
more fluctuating in the opening hours since that lots of news arrive
during the market closure and the market needs to rapidly response to
them at the opening hours. On the other hand, the investors tend to
make decisions after the market takes into account all information and
thus more transactions are made in the closing hours. To avoid this
daily oscillation, for the intraday data of each stock we divide the
trading value with its pattern $A(s)$.

\section{Scaling in the Trading Value and its Growth Rate}
 
{\it Scaling\/} and {\it universality} are two important concepts in
statistical physics. A system obeys a scaling law if its components
can be characterized by a property having a power-law relation for a
broad range of scales (``scale invariance''). A typical behavior for
scaling is {\it data collapse\/}: all curves can be ``collapsed'' onto
a single curve, after a certain scale transformation. In many systems,
the {\it same\/} scaling function holds, suggesting universal
laws. Now we are trying to test whether the market activity has
scaling and universality features.

\begin{figure*}
\begin{center}
   \includegraphics[width=\textwidth, angle = 0]{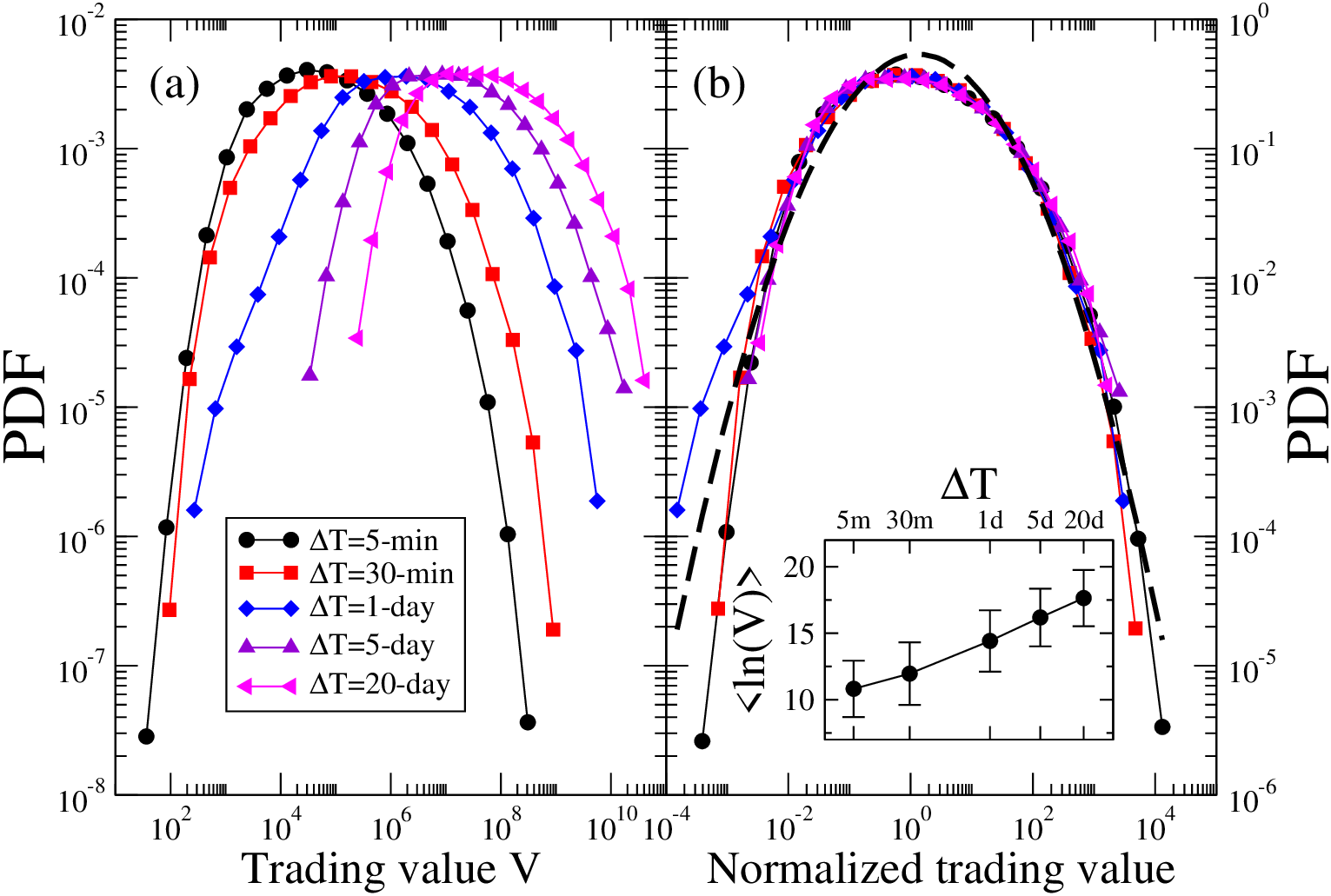}
\end{center}
\caption{(Color online) Distribution of the trading values. (a) Plot
of the probability density functions (PDF) of the trading value for 5
sampling intervals, $\Delta T= 5$-min, $30$-min, $1$-day, $5$-day, and
$20$-day. Though these curves have different centers and widths, their
shapes are similar. To test this similarity, (b) we normalize the
trading value for each sampling interval and plot their PDFs. All
curves almost collapse into a single one, which can be well-fit by a
log-normal distribution (as shown by the dashed line). This result
suggests that the trading value follows a log-normal distribution for
a sampling interval range over $3$ order of magnitudes. The mean value
of $ln(V)$, $\langle ln(V) \rangle$, vs. the sampling interval $\Delta
T$ is plotted in the inset of panel (b), where the error bars stand
for standard deviations of $ln(V)$, $\sigma(ln(V))$. Note that
$\langle ln(V) \rangle$ is logarithmically related to $\Delta T$ and
$\sigma(ln(V))$ is almost constant, suggesting that the trading value
linearly depends on the sampling interval. }
\label{Fig3}
\end{figure*}

We begin by examining the distribution of the trading values, which
can be characterized by a probability density function (PDF) $P$. In
Fig. \ref{Fig3}(a) we plot the PDFs of $5$ sampling intervals, $\Delta
T= 5$-min, $30$-min, $1$-day, $5$-day, and $20$-day for all the $3314$
stocks. The distributions shift to the right (large values of $V$)
with the increasing of the sampling interval. Interestingly, all the
five distributions have a similar shape, which approximately follows a
log-normal function,

\begin{equation}
P(V) \sim \exp{(-\frac{(ln(V)-\langle ln(V) \rangle)^2}{2 \sigma^2(ln(V))})}.
\label{P.eq}
\end{equation}
In this paper $\sigma(x) \equiv \sqrt{\langle x^2 \rangle - \langle x
\rangle^2}$ represents the standard deviation of variable $x$. To
further test Eq. (\ref{P.eq}), we normalize the trading value by
replacing $ln(V)$ with $(ln(V) -\langle ln(V) \rangle) /
\sigma(ln(V))$ and plot the corresponding PDFs in
Fig. \ref{Fig3}(b). Remarkably, all curves almost collapse onto a
single one. Moreover, these curves can be well-fit by a log-normal
function (as shown by the dashed line in Fig. \ref{Fig3}(b)). This
result supports that the distribution of the trading values follows
Eq. (\ref{P.eq}). We also plot the mean values and standard deviations
(as error bars) of $ln(V)$ in the inset of Fig. \ref{Fig3}(b). While
the mean value of $ln(V)$ increases with $\Delta T$, its standard
deviation is almost constant for all the sampling intervals. This
behavior further supports the consistency between the trading values
over a wide range of sampling intervals, from $5$ minutes to $20$
trading days.

Dynamics of the financial markets is a key issue in econophysics and
economics, which is usually characterized by the growth rate. This
measure has been well studied for the price, which smoothly evolve
with the time for many securities. However, the change in the equity
activity might be irregular. There might be few transactions in some
periods but many in the other periods. To avoid dramatic fluctuations,
we define the growth rate at time $t$, $g(t)$, as the logarithmic
change of two consecutive cumulative activities \cite{Rybski09}, i.e.,
for the trading value at time $t-1$, $V(t-1)$, and at time $t$,
$V(t)$,

\begin{equation}
g(t) \equiv ln(\frac{V(t-1) + V(t)}{V(t-1)}) = ln(1 + \frac{V(t)}{V(t-1)}).
\label{g.eq}
\end{equation}
We analyze the growth rate of all stocks at all times together,
therefore we neglect time $t$ of the growth rate. For simplicity, we
denote the initial trading value of the growth rate, $V(t-1)$, as
$V_i$ in the following. To examine the dynamics of the activity, we
study two measures of the growth rate: (i) the conditional mean growth
rate $\langle g|V_i \rangle$, which quantifies the average growth rate
of the trading value given the initial trading value $V_i$; (ii) the
conditional standard deviation $\sigma(g|V_i) \equiv \sqrt{\langle
g^2|V_i \rangle - \langle g|V_i \rangle^2}$, which characterizes the
fluctuation of the growth rate conditional on a given initial trading
value $V_i$.

\begin{figure*}
\begin{center}
   \includegraphics[width=\textwidth, angle = 0]{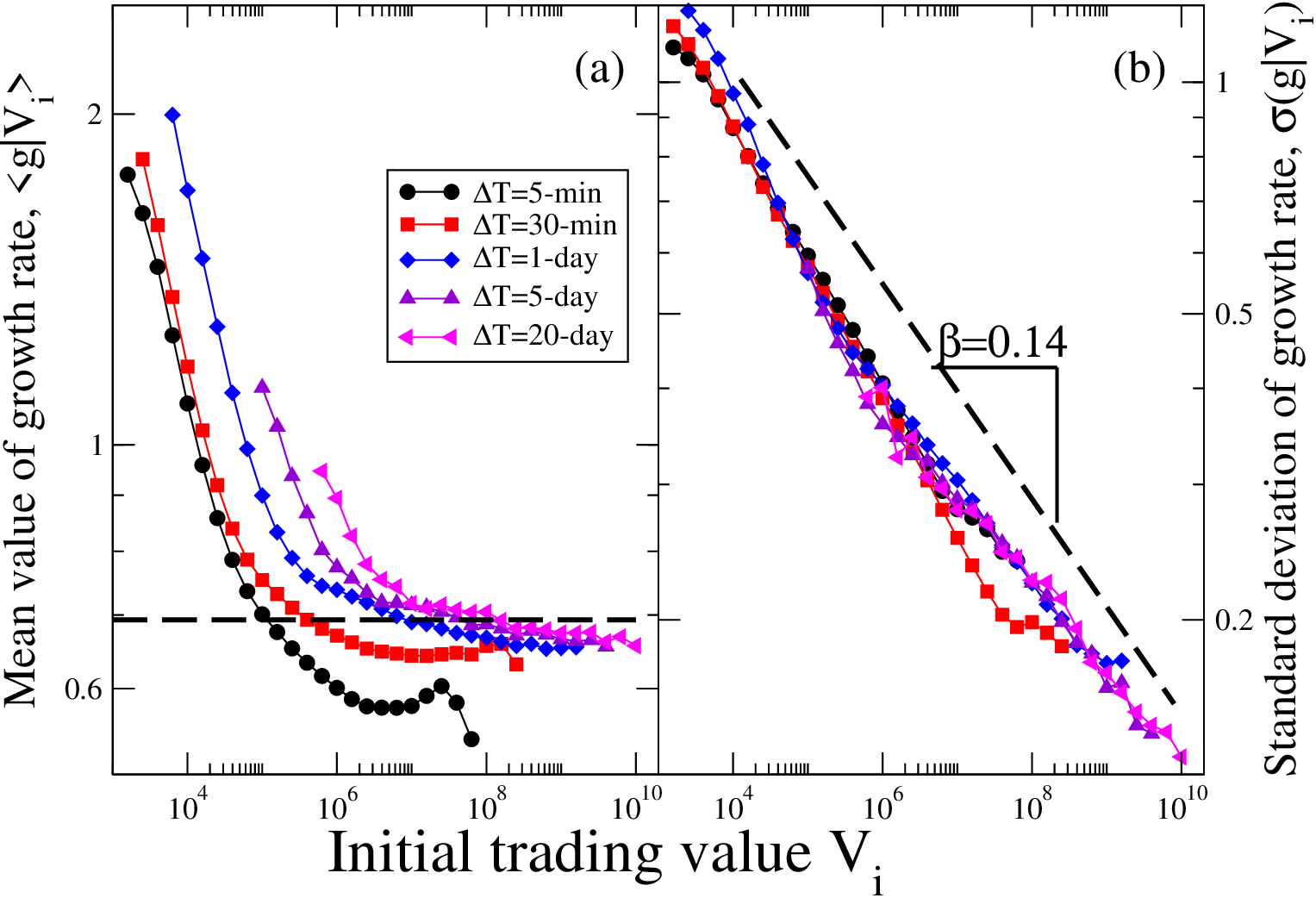}
\end{center}
\caption{(Color online) Scaling in the growth rate of the trading
value. The conditional mean growth rate $\langle g|V_i \rangle$ and
conditional standard deviation of the growth rate $\sigma(g|V_i)$
vs. the initial trading value $V_i$ are plotted in panel (a) and (b)
respectively. Five sampling intervals, $\Delta T= 5$-min, $30$-min,
$1$-day, $5$-day, and $20$-day, are explored. Remarkably, al curves in
panel (a) converge to a constant value, $ln(2)$, as suggested by the
dashed line. Thus the two consecutive trading values tend to be the
same, which indicates a strong memory in the sequence of the trading
value. Note that all curves have large $\langle g|V_i \rangle$ values
for small $V_i$ values, and they shift to the right when the sampling
interval is increased. Also note that the curves of $\Delta T=5$-min
and $30$-min have small drops for large $V_i$ values. Both behaviors
are due to the finite size effect in the data. In panel (b), all
curves approximately collapse into a single one, which suggests a
scaling law for the sampling interval over $3$ order of
magnitudes. Moreover, the curves can be well-fit by a power law for
the trading value over $6$ order of magnitudes, $\sigma \sim
V_i^{-\beta}$ with exponent $\beta=0.14 \pm 0.02$. }
\label{Fig4}
\end{figure*}

In Fig. \ref{Fig4}(a) we plot the conditional mean value $\langle
g|V_i \rangle$ vs. initial value $V_i$ for the five sampling
intervals. Interestingly, all the five curves tend to a constant
$ln(2)$ for large $V_i$ values (as shown by the dashed
line). According to Eq. (\ref{g.eq}), this behavior suggests that two
consecutive trading values tend to be similar for the whole market. In
other words, there is certain memory in the time series of trading
value. In addition, there is an obviously systematic tendency with the
sampling intervals. With the increasing of $\Delta T$, $\langle g|V_i
\rangle$ decreases and shifts to the right for small $V_i$ values,
which is due to the limited size of the database. For small $V_i$
values, the next trading value is typical the same or larger, thus the
corresponding $\langle g|V_i \rangle$ value is significantly larger
than $ln(2)$. For very large $V_i$ values, the next trading values
tend to be smaller and thus $\langle g|V_i \rangle$ values decrease,
as shown in the curves of $\Delta T=5$-min and $30$-min.

Next we plot the conditional standard deviation $\sigma(g|V_i)$
vs. $V_i$ in Fig. \ref{Fig4}(b), and find that the curves for the all
five sampling intervals collapse onto a single one. Furthermore, these
curves follow a power-law function (as guided by the dashed line in
Fig. \ref{Fig4}(b)),

\begin{equation}
\sigma(g|V_i) \sim V_i^{-\beta},
\label{beta.eq}
\end{equation}
with the exponent $\beta=0.14 \pm 0.02$, which is consistent with the
finding on growth rate of other complex systems
\cite{Stanley96,Rybski09}. We must note that this scaling persists for
a very broad range of $V_i$ values, which covers more than $6$ order
of magnitudes. This remarkably behavior indicates a significant
universality in the whole equity market.

\section{Long-Term Correlations}

Many financial time series have memory, where a value in the sequence
depends on the previous values. Previous studies have shown that the
return does not exhibit any linear correlations extending for more
than a few minutes, but the volatility exhibits long-term correlations
(see Refs. \cite{Mantegna00} and \cite{Liu99} for example). Thus, the
temporal structure in the equity activity is also of
interest. Fig. \ref{Fig4}(a) already suggests a certain memory between
two consecutive values of $V$. To further test the correlations in the
trading value time series, we employ the detrended fluctuation
analysis (DFA), a wide-used method to examine the correlations in the
time series
\cite{Peng94,Peng95,Bunde00,Hu01,Chen02,Kantelhardt02,Xu05}. In
contrast to the conventional method such as the auto-correlation
function, DFA can deals with the non-stationary time series such as
financial markets records. After removing trends, DFA computes the
root-mean-square fluctuation $F(\ell)$ of a time series within a
window of $\ell$ points, and determines the Hurst exponent $H$ from
the scaling function,
\begin{equation}
F(\ell)\sim\ell^H.
\label{dfa.eq}
\end{equation}
The correlation is characterized by the Hurst exponent $H\in(0,1)$. If
$H>0.5$, the records have positive long-term correlations. if $H=0.5$,
no correlation (white noise), and if $H<0.5$, it has long-term
anti-correlations.

\begin{figure*}
\begin{center}
   \includegraphics[width=\textwidth, angle = 0]{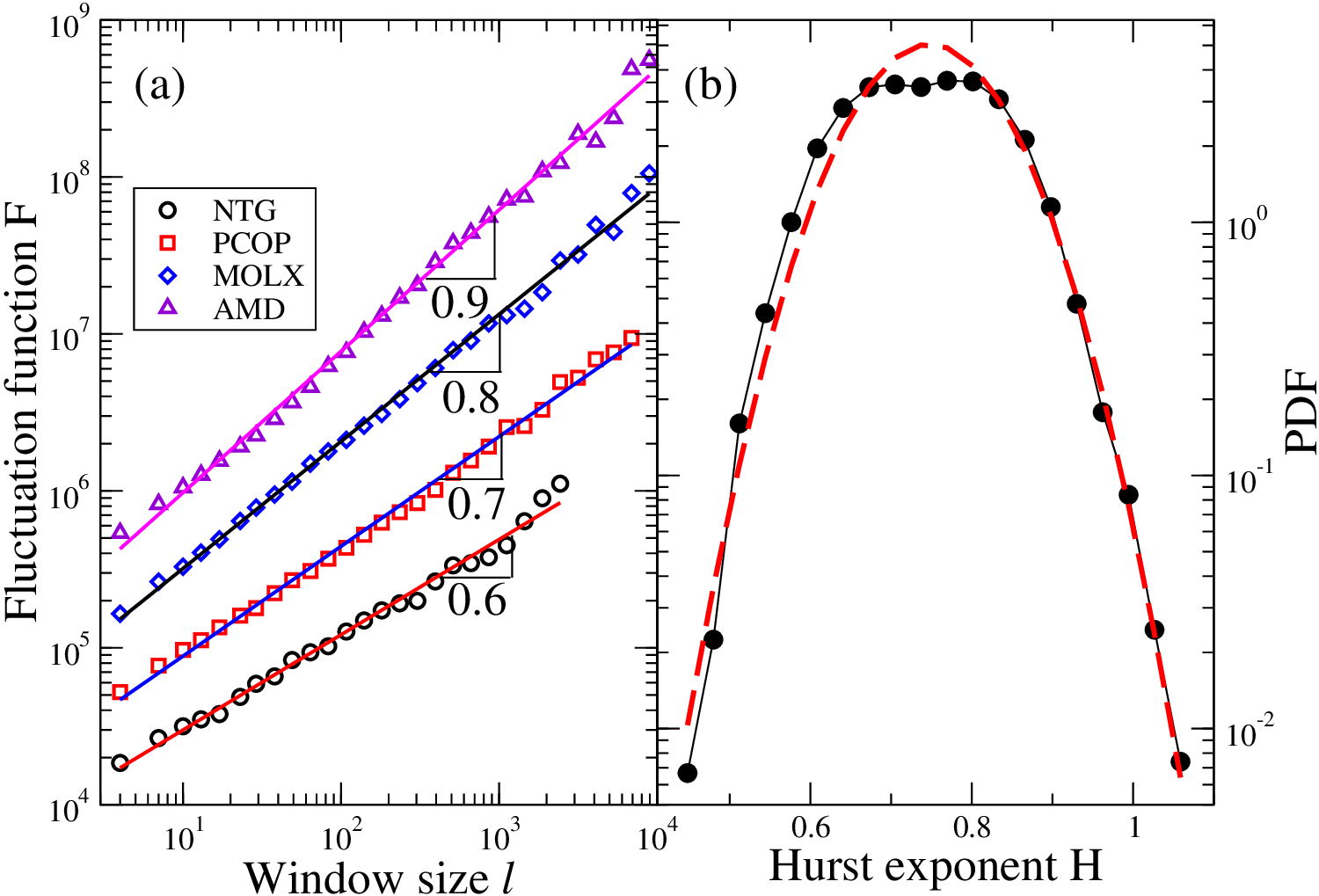}
\end{center}
\caption{(Color online) Long-term correlation in the time series of
the trading value. The time series is for a typical sampling interval,
$\Delta T=5$-min. (a) Illustration of the DFA plot, fluctuation
function $F$ vs. window size $\ell$.  The results of four
representative stocks, NTG, PCOP, MOLX, and AMD, are plotted and their
Hurst exponents $H$ are labeled in the panel. (b) Distribution of the
Hurst exponent $H$ for the 3314 stocks. Note that the distribution can
be well-fit by a normal distribution, as shown by the dashed line,
with $\langle H \rangle=0.75$ and $\sigma(H)=0.09$.}
\label{Fig5}
\end{figure*}

Without loss of generality, we study the long-term correlation in the
time series of $\Delta T=5$-min for every stock. As examples, we plot
the DFA curves in Fig. \ref{Fig5}(a) for the activity of four typical
stocks, Natco Group Inc. (NTG), Pharmacopeia Drug Discovery
Inc. (PCOP), Molex Inc. (MOLX), and Advanced Micro Devices
Inc. (AMD). As seen in the plot, their corresponding Hurst exponent
varies in a wide range, from $0.6$ to $0.9$. To investigate the
long-term correlations for all the $3314$ stocks, we plot the
distribution of the $3314$ $H$ values in
Fig. \ref{Fig5}(b). Interestingly, this distribution can be well
characterized by a normal distribution with a mean value of $0.75$ and
standard deviation of $0.09$, as seen by the dashed line in the plot.

\begin{figure*}
\begin{center}
   \includegraphics[width=\textwidth, angle = 0]{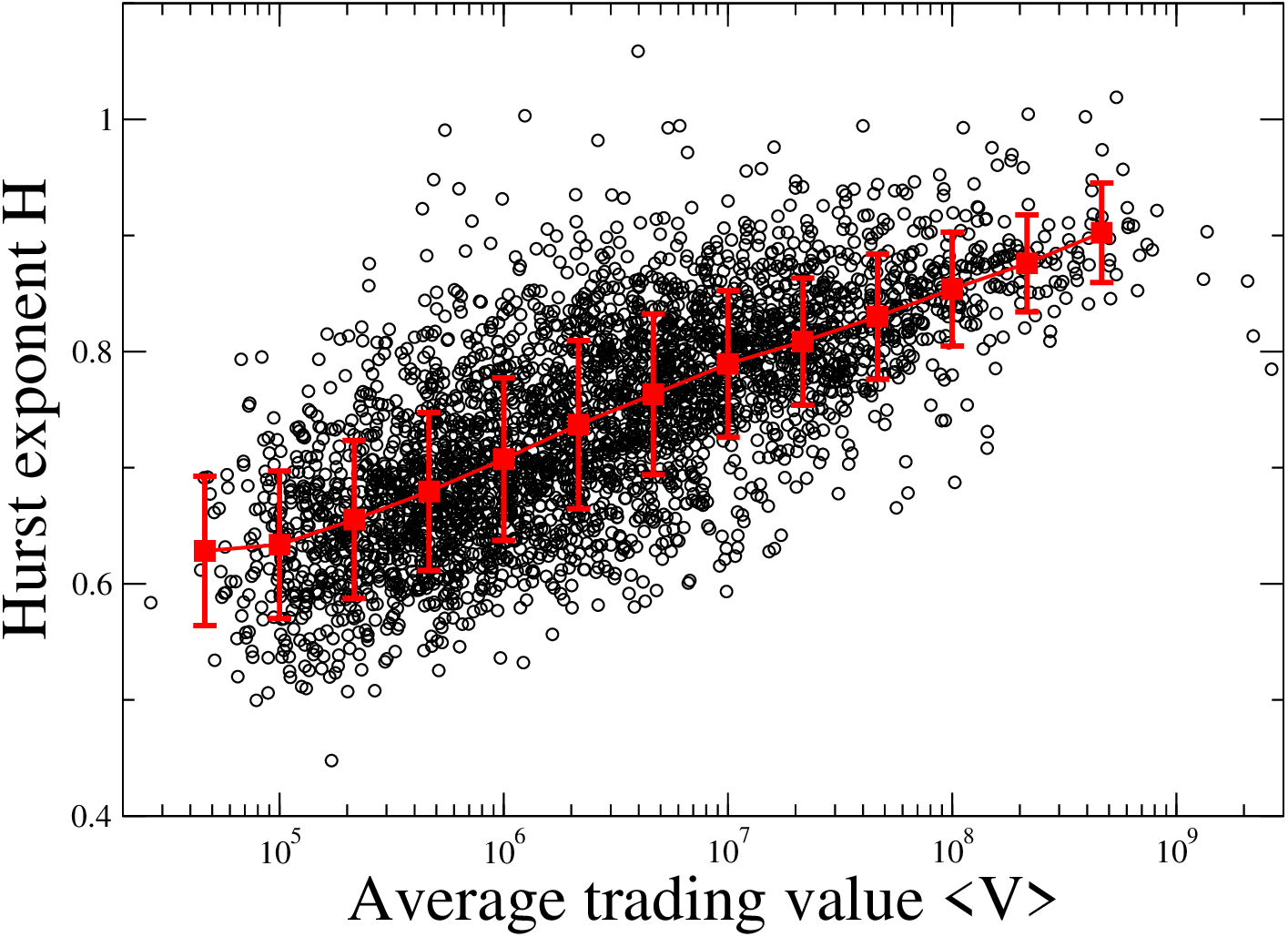}
\end{center}
\caption{(Color online) Scatter plot of Hurst exponent $H$ vs. average
trading value $\langle V \rangle$ for the $3314$ stocks. To better
show the tendency, we group the points with similar $\langle V
\rangle$ values and plot their mean values (as shown by the filled
squares) and standard deviations (as shown by the error bars). It is
seen that $H$ depends on the size of $\langle V \rangle$, and their
relation approximately follows a logarithmic function, $H \sim 0.033
\times ln(\langle V \rangle)$. This finding suggests that the stocks
of smaller size are less correlated. It seems that their activities
are more influenced by the external factors. }
\label{Fig6}
\end{figure*}

The next question is, why the Hurst exponent $H$ varies with the
stock? As we know, there are many factors that relate to the long-term
correlations in the financial time series. For example, the long-term
correlations in the volatility depends on the size, activity, risk,
and return of the stock \cite{Wang09}. Thus there might be many
factors that affect the long-term correlations in the activity. Here
we test the relation between $H$ and the average trading value
$\langle V \rangle$ (without loss of generality, we choose the
sampling interval of $\Delta T=1$-day for $\langle V \rangle$). The
exponent $H$ clearly shows dependence on the value of $\langle V
\rangle$, as seen in the scatter plot of Fig. \ref{Fig6}. The exponent
$H$ of activity tends to increase with $\langle V \rangle$. To better
show the tendency, we also plot the average and standard deviation (as
the error bar) of $H$ values for every logarithmic bin of $\langle V
\rangle$ values, which follows a logarithmic function, $H \sim 0.033
\times ln(\langle V \rangle)$. This finding indicates that the
long-term persistence is relative weaker for the stocks with smaller
$\langle V \rangle$ values. This might be understood since small
stocks are easier to be influenced by the external factors and events
due to their small size and market depth. Note that the error bars are
almost constant for all range of $\langle V \rangle$ values.

\section{Discussions}

As discussed in the Section II, other equity activity measures such as
the number of trades and trading volume strongly depends on the
trading value. We also explore their features including the scaling of
the distributions and the long-term correlations in the activity time
series. As expected, they are similar to those obtained for the
trading value.

As suggested by Rybski et al. \cite{Rybski09}, the exponent $\beta$
and $H$ are related to each other and represent the fluctuations in
the time series. Using Eqs. (\ref{beta.eq}) and (\ref{dfa.eq}),
$\beta$ is determined by the fluctuation of growth and its scaling
with the size of the initial activity, and $H$ is generated from the
scaling of the fluctuation with the time interval. This leads to the
relation between $\beta$ and $H$ \cite{Rybski09}

\begin{equation}
\beta = 1 - H.
\label{relation.eq}
\end{equation}
In our analysis we find $\beta = 0.14$ (Fig. \ref{Fig4}(b)) and $H$
close to $0.8 $ for all the $3314$ stocks (Fig. \ref{Fig5}(b)), which
roughly satisfies Eq. (\ref{relation.eq}). Fig. \ref{Fig4}(b)
accumulates all the data points of the $3314$ stocks while the
correlations vary with the stock, as demonstrated by
Fig. \ref{Fig5}(b). To better understand the dynamics of equity
activity, the relation between $\beta$ and $H$ should be
comprehensively examined, which will be studied in the future.

In summary, we studied the activity of the $3314$ most-traded
U.S. stocks. We showed that the equity activity has an intraday
pattern. The stock is more traded in the opening hours and
significantly more in the closing hours. We found that the
distribution of activity follows a log-normal function for all studied
sampling intervals. We also found that the conditional standard
deviation of growth rate has a power-law dependence on the initial
trading value. Moreover, this scaling behavior is persistent for a
wide range of sampling intervals, from $5$ minutes to $20$ trading
days. Further, we explored the long-term correlations in the time
series of the equity activity and found that the correlation exponent
$H$ has a normal distribution over the entire market with
$H=0.75\pm0.09$. We also showed that the long-term correlation
exponent $H$ depends logarithmically on the size of the activity. In
other words, the smaller stock tends to have weaker long-term
correlations.

\section*{Acknowledgments}

We thank the NSF and Merck Foundation for financial support.

\newpage

\end{document}